\documentclass[a4paper]{article}

\usepackage{INTERSPEECH2020}

\usepackage{float}
\usepackage{graphicx}
\usepackage{caption}
\usepackage{subcaption}
\usepackage{arydshln}
\usepackage{mathtools}
\DeclarePairedDelimiter\floor{\lfloor}{\rfloor}
\usepackage{amsmath}

\usepackage{siunitx}				
\newcommand{\PreserveBackslash}[1]{\let\temp=\\#1\let\\=\temp}
\newcolumntype{C}[1]{>{\PreserveBackslash\centering}p{#1}}
\newcolumntype{R}[1]{>{\PreserveBackslash\raggedleft}p{#1}}
\newcolumntype{L}[1]{>{\PreserveBackslash\raggedright}p{#1}}
\newcommand\Tstrut{\rule{0pt}{2.6ex}}         
  
\usepackage{booktabs}

\usepackage{multirow}
\usepackage{multicol}

\title{Improvement of Noise-Robust Single-Channel Voice Activity Detection with Spatial Pre-processing}
\name{Max Væhrens$^1$, Andreas Jonas Fuglsig$^2$, Anders Post Jacobsen$^3$, Nicolai Almskou Rasmussen$^4$, Victor Mølbach Nissen$^5$, Joachim Roland Hejslet$^6$, Zheng-Hua Tan$^7$}
\address{
  $^{1,2,4,5,6,7}$Department of Electronic Systems, Aalborg University, Aalborg, Denmark\\
  $^{2,3}$RTX A/S, Nørresundby, Denmark
  }
\email{$^1$maxwaehrens@gmail.com, $^2$ajf@rtx.dk, $^3$apj@rtx.dk, $^4$nara17@student.aau.dk, $^5$vnisse17@student.aau.dk, $^6$jhejsl20@student.aau.dk, $^7$zt@es.aau.dk}

\begin{document}

\maketitle
\begin{abstract}
Voice activity detection (VAD) remains a challenge in noisy environments.
With access to multiple microphones, prior studies have attempted to improve the noise robustness of VAD by creating multi-channel VAD (MVAD) methods.
However, MVAD is relatively new compared to single-channel VAD (SVAD), which has been thoroughly developed in the past.
It might therefore be advantageous to improve SVAD methods with pre-processing to obtain superior VAD, which is under-explored.
This paper improves SVAD through two pre-processing methods, a beamformer and a spatial target speaker detector.
The spatial detector sets signal frames to zero when no potential speaker is present within a target direction.
The detector may be implemented as a filter, meaning the input signal for the SVAD is filtered according to the detector's output; or it may be implemented as a spatial VAD to be combined with the SVAD output.
The evaluation is made on a noisy reverberant speech database, with clean speech from the Aurora 2 database and with white and babble noise.
The results show that SVAD algorithms are significantly improved by the presented pre-processing methods, especially the spatial detector, across all signal-to-noise ratios.
The SVAD algorithms with pre-processing significantly outperform a baseline MVAD in challenging noise conditions.
\end{abstract}

\section{Introduction}
Voice activity detection (VAD) has come a long way since its origin, but high noise environments remain a significant challenge.
To improve performance in very noisy conditions, previous studies have utilized multiple microphones to successfully develop multi-channel VADs (MVADs) that base the VAD decision on spatial cues, e.g. interchannel time, level and phase difference (ITD, ILD and IPD respectively) \cite{ILD} \cite{dualMicReliableSpatialCues} \cite{vad_distant_talking}.
However, MVADs are relatively new compared to single-channel VAD (SVAD) algorithms that have been developed very thoroughly \cite[p. 2]{rVAD}.
Therefore, it might be advantageous to improve the existing SVAD by applying pre-processing techniques to the algorithms.
To the best of the authors' knowledge, prior studies have not investigated the use of pre-processing methods based on spatial cues to improve multiple existing SVAD algorithms.

In this paper we propose to use pre-processing methods based on spatial cues for improvement of SVAD algorithms.
Two pre-processing methods are presented: The first consists of beamforming and the second a spatial target speaker detector, which is used to set signal frames to zero when there are only signals from an undesired direction.
The latter method can be combined with SVAD in two different approaches: The first approach filters the input to SVAD algorithms while the second approach is a spatial VAD which is combined with the SVAD decision.
The beamformer and spatial detector methods may also be combined, which in total yields four types of pre-processing approaches.

We consider three SVAD algorithms: the unsupervised segment-based robust VAD (rVAD) by Tan et al. \cite{rVAD}, \cite{tan2010low}; the ITU-T recommendation G.729 annex B voice encoder (G729B) \cite{G729_annexB}; and the statistical model-based VAD (SOHN) by Sohn et al. \cite{statistical_model-based_vad}.
The combination of the proposed pre-processing methods and SVAD algorithms is compared with a MVAD called frequency selective normalised difference power spectral density (FS-NDPSD) VAD by S. Hwang et al. \cite{dualMicReliableSpatialCues}.
To evaluate the proposed method acoustic environments are simulated.
The results show that all of the presented SVAD algorithms can be improved by the pre-processing methods.
Furthermore, for low signal-to-noise ratios SVAD with pre-processing outperforms the FS-NDPSD.

\section{Spatial pre-processing methods}\label{sec:improving_svad}
The proposed method is to improve SVAD performance by combining a dual-channel pre-processor with SVAD.
The two sampled microphone signals $s(k)$ are given by:
\begin{align}\label{eq:signal_model}
    s_1(k) &= a_1(k)\cdot x(k)+n_1(k), \\
    s_2(k) &= a_2(k)\cdot x(k+\tau)+n_2(k),
\end{align}
where, $k$ is the sample index; $x(k)$ is the speech signal; $a_i$ is the attenuation of the speech signal for each microphone; $n_i(k)$ is the noise at each microphone; and $\tau$ is the ITD.
For implementation, time frames are utilised to achieve quasi-stationary speech \cite[p. 16-20]{discrete-time_processing_of_speech_signals}, and time frames are indexed by $t$.

\begin{figure}
\centering
\subfloat[Spatial filter method with an optional beamformer (dashed line) \label{subfig:fb-svad}]{\includegraphics[width=1\linewidth]{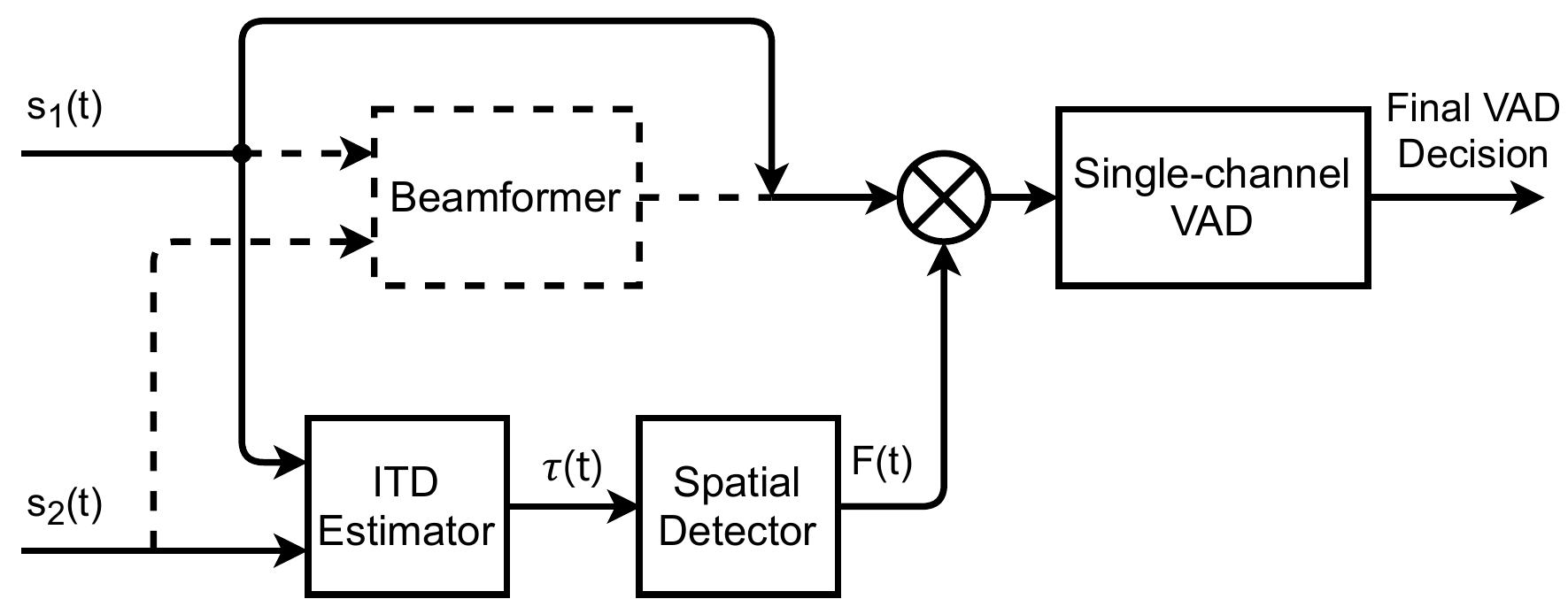}} \,
\subfloat[Spatial VAD method with an optional beamformer (dashed line) \label{subfig:ab-svad}]{\includegraphics[width=1\linewidth]{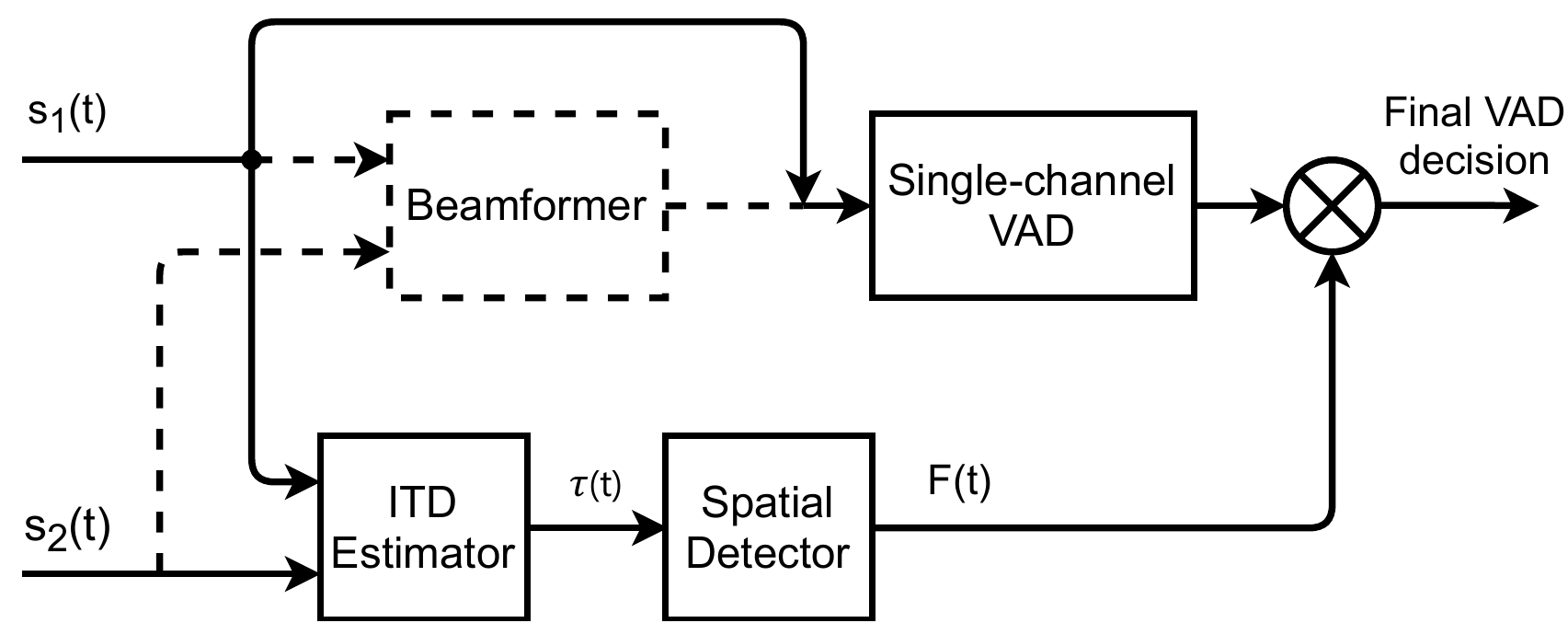}}
\captionsetup{belowskip=-15pt}
\caption{Flowchart of the proposed method for the two combination approaches}
\label{fig:flowchart}
\end{figure}

The spatial target detector is the first part of the proposed method, this is described in Section~\ref{subsec:SpatialFilter}.
The detector may be combined with a SVAD algorithm in two approaches, as is seen in the two block diagrams in Figure~\ref{fig:flowchart}, and these two approaches are described in Section~\ref{subsec:TwoApproaches}.
Beamforming is the second pre-processing method and it may be combined with the spatial detector, this is illustrated by the dashed lines in Figure~\ref{fig:flowchart} and it is described in Section~\ref{subsec:Beamformer}.

\subsection{Spatial target detector}\label{subsec:SpatialFilter}
\begin{figure}
\captionsetup{belowskip=-15pt}
    \centering
    \includegraphics[width=\linewidth]{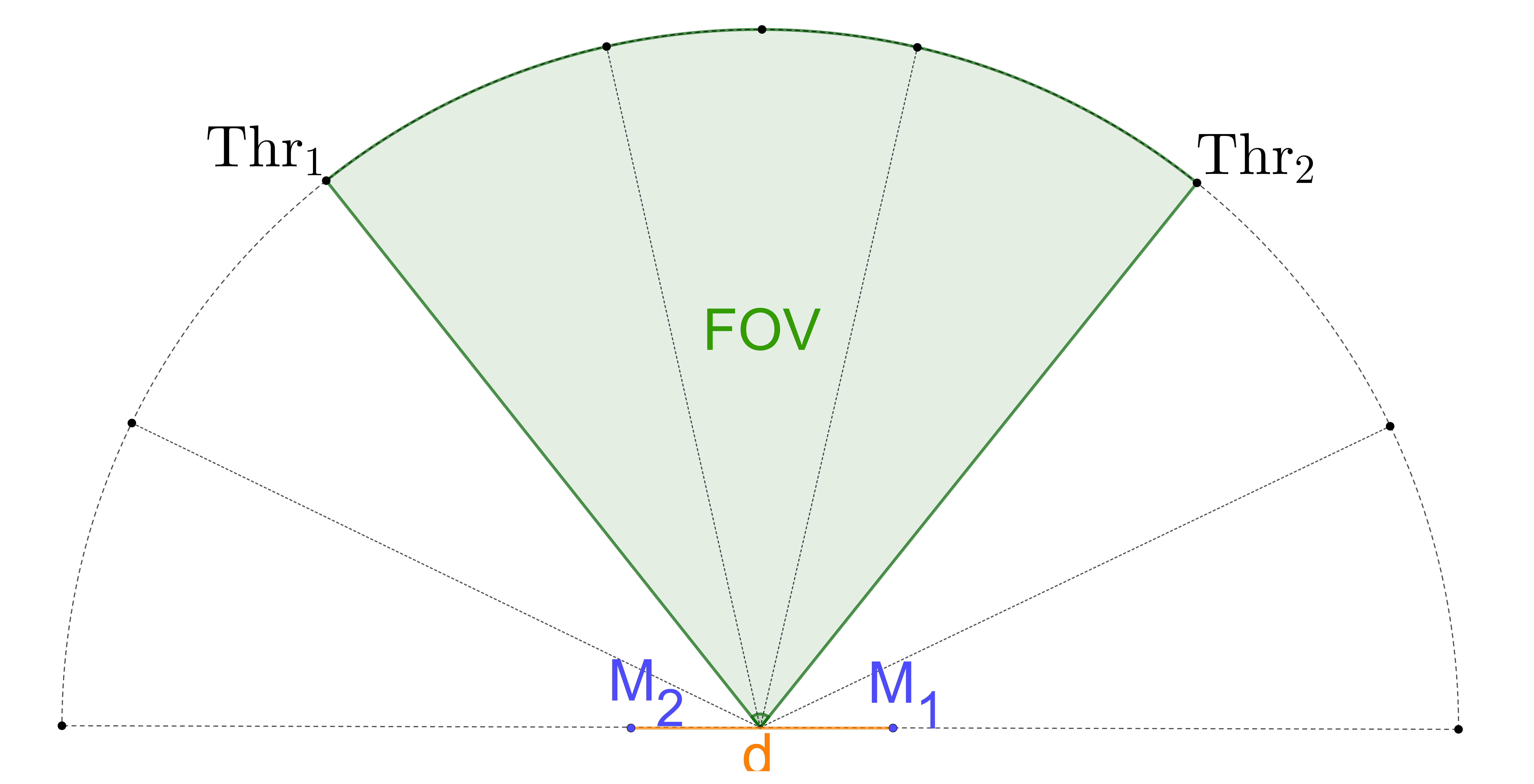}
    \caption{Overview of the dual microphone array, thresholds and FOV}
    \label{fig:fov}
\end{figure}

The spatial target speaker detector works by indicating whether a target speaker is within a target direction for a given time frame.
A potential speaker is detected if the measured ITD for a given time frame, $\tau(t)$, of the two microphones is within two thresholds, $\text{Thr}_1$ and $\text{Thr}_2$, as seen in \eqref{eq:filter_threshold}:
\begin{equation}\label{eq:filter_threshold}
    F(t) = \begin{cases}\displaystyle1,\qquad\text{if}\quad  \text{Thr}_1 \le \tau(t) \le \text{Thr}_2, \\ 0, \qquad \text{otherwise}.\end{cases}
\end{equation}
Spatially, the thresholds will create a field of view (FOV), which consists of the directions of arrival for the given application where a potential target is expected.
The FOV and arbitrary thresholds are illustrated on Figure~\ref{fig:fov} along with a microphone array, M$_1$ and M$_2$.

The maximum ITD, $\tau_{\mathrm{max}}$, for a given application is determined by the geometry of the array, $\mathrm{d}$, and the sampling frequency, $f_\mathrm{s}$:
\begin{equation}\label{eq:maxITD}
    \tau_{\mathrm{max}}=\floor*{\frac{f_\mathrm{s}}{\mathrm{c}/\mathrm{d}}} \cdot 2 + 1,
\end{equation}
where $\mathrm{c}$ is the speed of sound in air.
Then, \SI{180}{\degree} may be divided by $\tau_{max}$ to obtain the angle resolution for the ITD, as is indicated by the slices of the half circle in Figure~\ref{fig:fov}.
In turn, the two thresholds for the FOV can be set to match the expected directions of arrival for a target speaker.
The thresholds may be updated adaptively, i.e. panning towards the target.

In this paper, the Generalized Cross Correlation with Phase Transformation (GCC-PHAT) is the chosen ITD estimation method \cite{GCC_phat}.
GCC-PHAT is chosen because it has good and efficient performance for speech applications, especially in reverberant environments \cite[p. 2]{Performance_optimization_GCCPHAT} \cite[p. 35]{gcc_comparison}.
\subsection{Combining the proposed spatial detector and SVAD}\label{subsec:TwoApproaches}
The spatial detector in~\eqref{eq:filter_threshold} can be combined with a SVAD in two different approaches.
In the first approach the detector output, $F(t)$, is used to filter the input signal, $s_1(t)$, which is sent to the SVAD, as seen in Figure~\ref{subfig:fb-svad}, and this is denoted F-SVAD.
When $F(t)$ is equal to one $s_1(t)$ is left unchanged, and when $F(t)$ is zero $s_1(t)$ is set to zero.
In the second approach~\eqref{eq:filter_threshold} is used as a spatial VAD.
The spatial detector output, $F(t)$, is combined with the SVAD decision in an AND operation, as seen in Figure~\ref{subfig:ab-svad}, and this is denoted A-SVAD.
Due to the place where the detector output and the SVAD decision are combined, these two can be computed in parallel.
\subsection{Combining the spatial filter with beamforming}\label{subsec:Beamformer}
An important feature of the proposed detector method is that both approaches can be combined with existing beamforming techniques.
This feature ensures that for some applications the performance can be even further improved.
This is also shown in Figure~\ref{fig:flowchart}, in both diagrams the dashed lines denote the beamformer and the solid line above the beamformer box is omitted when the beamformer is used.
Here the F-SVAD and A-SVAD combined with the beamformer becomes FB-SVAD and AB-SVAD, respectively.
For this paper, a delay and sum (DS) beamformer \cite{Delay_and_sum} is chosen to show the performance increase obtained by including additional beamforming.
Other beamforming methods such as minimum variance distortionless response, linearly constrained minimum variance and multi-channel Wiener filter may substitute the DS beamformer to obtain even better results \cite{multichannelTechniques}.

\section{Evaluation}\label{sec:evaluation}
Objective evaluation of the SVAD and SVAD with pre-processing is made on a simulated noisy reverberant speech database. 
The clean speech utterances are from the Aurora 2 database \cite[p. 2]{Aurora-2_database}.
The custom database is simulated with the image-source method (ISM), specifically the MATLAB implementation by Lehmann and Johansson is used \cite{ISM_eric_lehmann_2008} \cite{ISM_eric_lehmann_web}.
\\
For the dual microphone array, the spacing of the microphones, $\mathrm{d}$, is \SI{15}{\centi\meter}.
$f_\mathrm{s}$ is limited by the Aurora 2 database at \SI{8}{\kilo\hertz}, thus $\tau_{\mathrm{max}}=7$ according to~\eqref{eq:maxITD}, which gives an angle resolution of \SI{25}{\degree}.
The thresholds in this implementation are set to $\text{Thr}_1=-1$ and $\text{Thr}_2 = 1$, i.e. three samples in total, which gives a FOV of \SI{75}{\degree}.
The frame shift is set to \SI{10}{\milli\second}.
The rVAD, G729B and SOHN VAD implementations which are used for the evaluation are the MATLAB code from \cite{zheng_hua_web}, \cite{g729_mathworks} and \cite{mike_brookes_voicebox} respectively.

\subsection{ISM simulation specifications}\label{subsubsec:simulation_specs}
The room size is specified according to an exemplary office room which is \SI{9.5 x 6.5 x 5}{\meter} (length $\times$ width $\times$ height) \cite[p. 1075]{dualMicVadTwoStepPowerLevel}.
The target speaker source is placed \SI{\sim0.39}{\meter} away from the center of the array.
The two microphones and the source are positioned at coordinates $(x, y, z)$ which are $\left(\SI{4.825}{\meter}, \SI{3.25}{\meter}, \SI{1.7}{\meter}\right)$, $\left(\SI{4.675}{\meter}, \SI{3.25}{\meter}, \SI{1.7}{\meter}\right)$ and $\left(\SI{4.75}{\meter}, \SI{2.857}{\meter}, \SI{1.7}{\meter}\right)$ respectively.
Finally, the reverberation time of the room $T_{60}$ is set to \SI{0.2}{\second} and all six surfaces of the room are set to have equal reflection coefficient.

\subsection{Simulating reverberant noise}\label{subsec:noise_simulation}
To simulate reverberant noise, the noise sources are placed within the acoustic environment by the ISM.
This gives a number of RIRs for the noise sources which each may be convolved with noisy sounds, thus placing the noisy sounds within the room.
Recorded noise such as the noise files from the Aurora 2 database \cite{Aurora-2_database} cannot be placed in this way, because the recordings contain different Room Impulse Responses (RIRs) compared to the noise sources in the simulated room, as discussed in \cite{ace_corpus_article}.
However, clean speech utterances can be placed at each noise source location to simulate competing speakers.
Utterances from the NOIZEUS database \cite{subjective_comparison_and_evaluation_of_speech} \cite{noizeus_database} are placed in this way at six different locations in the room. 
The six locations are placed \SI{3}{\meter} away from the center point of the microphone array in a circle with a height fixed at \SI{1.7}{\meter}.

The speech and noise are then summed at different SNRs \cite[p. 2]{Aurora-2_database}, to create various levels of noise scenarios.
The SNR is defined as the ratio of active speech energy compared to the constant noise energy.
Active speech energy is determined using the ITU-T P.56 recommendation \cite{itu_p56}.
The noise energy is the root mean square value across the whole noise segment.
According to the measured energy of speech and noise, the noise is scaled appropriately to obtain four SNR levels (\SIlist{-5;0;10;20}{\decibel}) before summing the clean speech and noise.
\subsection{Objective evaluation}
Reference labels can accurately be made with a forced-alignment speech recognition algorithm \cite[p. 2940]{comparison_referenceVAD}.
Accordingly, labels are made for the Aurora 2 database found at \cite{zheng_hua_web}.
The VAD algorithms are evaluated by comparing the frame-level VAD results with the reference labels for every speech segment in the test set.
Two measurements are then made for the evaluated VAD, one for the number of correctly detected speech frames (speech detection rate (SDR)) and one for the number of non-speech frames detected as speech (false acceptance rate (FAR)).
Based on these two statistics, the receiver operating characteristics (ROC) curve is plotted.
To numerically evaluate the VAD performance the area under the ROC curve (AUC) is calculated \cite[p. 89]{detection_theory}.
\section{Results} \label{sec:results}
The purpose of the first evaluation is to determine which approach of the spatial detector (F-SVAD and A-SVAD) is the best for each of the SVAD algorithms.
The best approach is then used in the second evaluation.
The purpose of the second evaluation is to compare the SVAD algorithms against the SVAD with the different pre-processing methods.
Additionally, the FS-NDPSD is also used as a baseline MVAD for comparison.
\subsection{Evaluation of the spatial detector methods combined with SVAD}
\begin{figure}[t]
\centering
\subfloat{\includegraphics[width=1\linewidth]{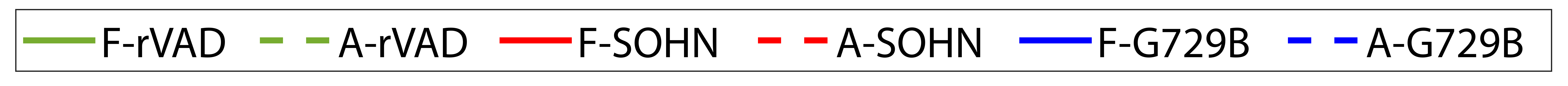}} \,
\setcounter{subfigure}{0}
\subfloat[\SI{10}{\decibel} white noise\label{subfig:results_combine_a}]{\includegraphics[width=0.5\linewidth]{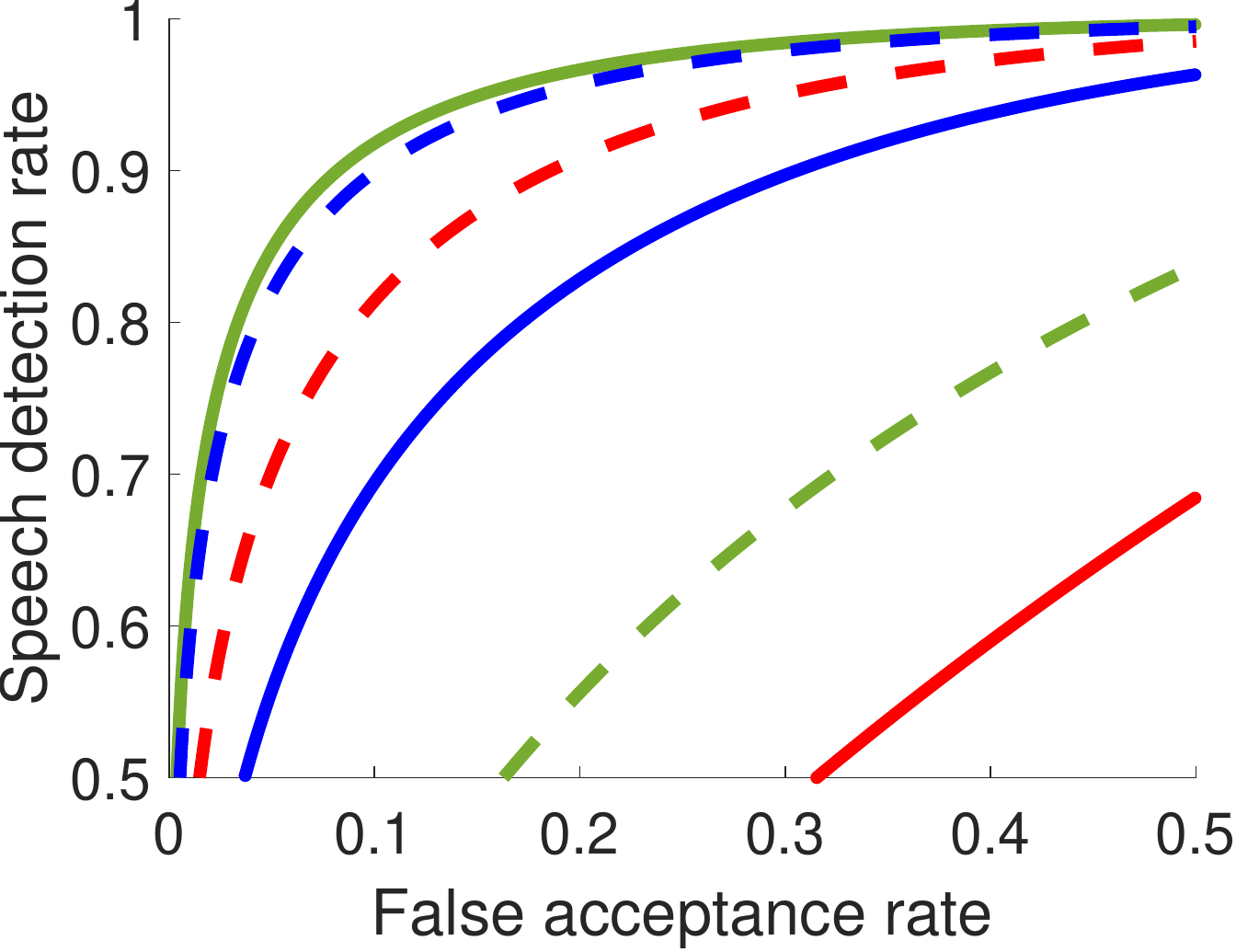}}
\subfloat[\SI{0}{\decibel} white noise\label{subfig:results_combine_b}]{\includegraphics[width=0.5\linewidth]{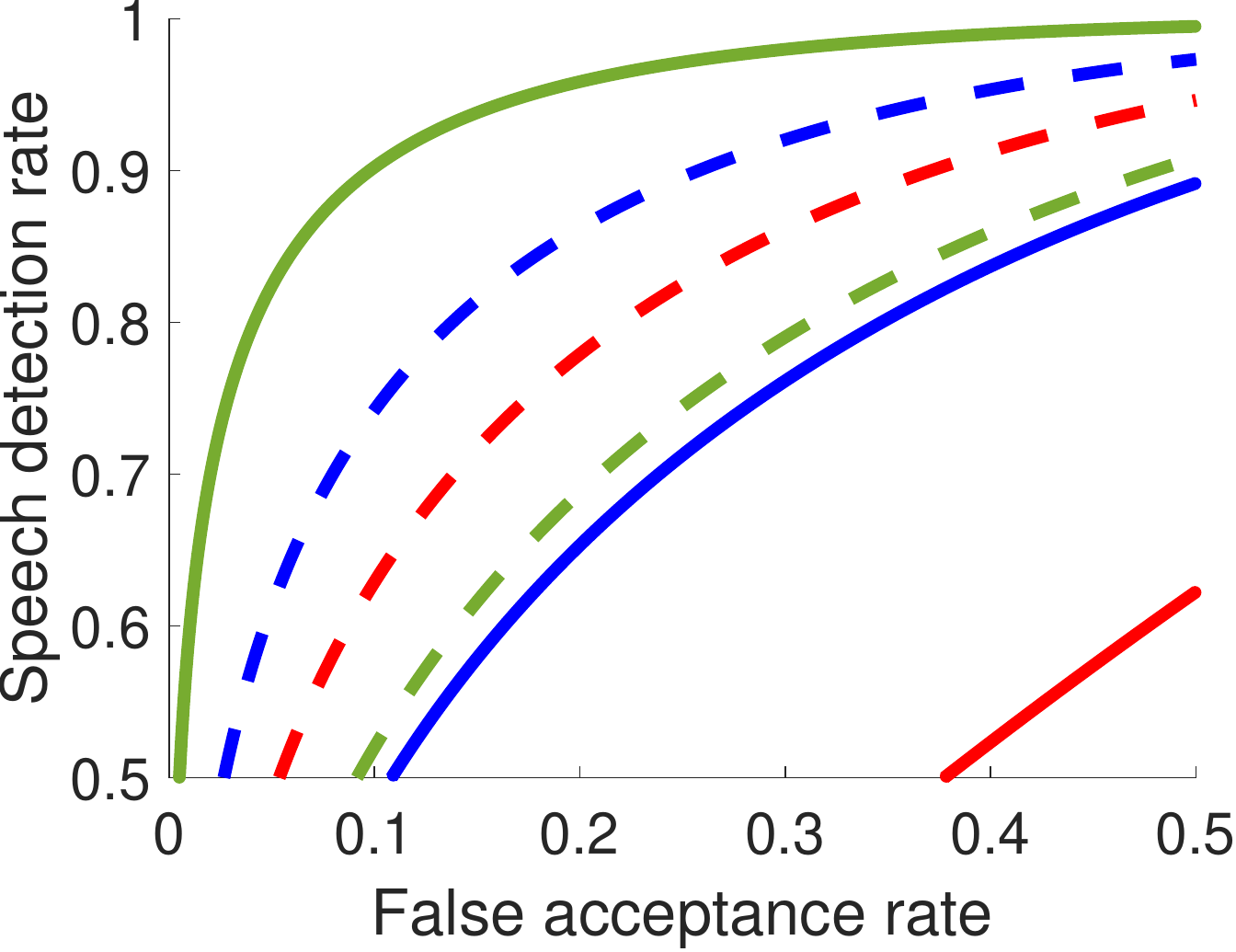}}
\,
\subfloat[\SI{10}{\decibel} babble noise\label{subfig:results_combine_c}]{\includegraphics[width=0.5\linewidth]{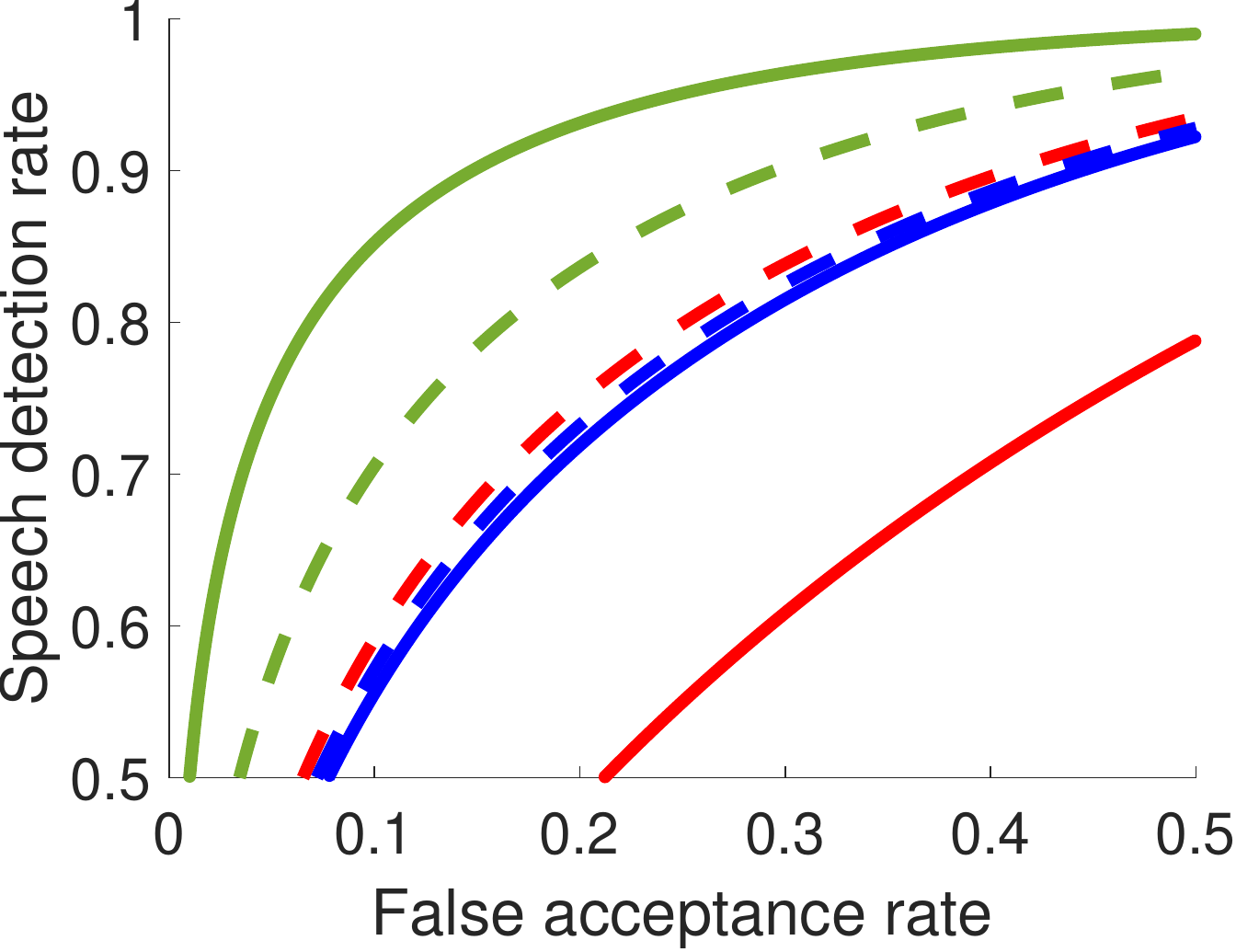}}
\subfloat[\SI{0}{\decibel} babble noise\label{subfig:results_combine_d}]{\includegraphics[width=0.5\linewidth]{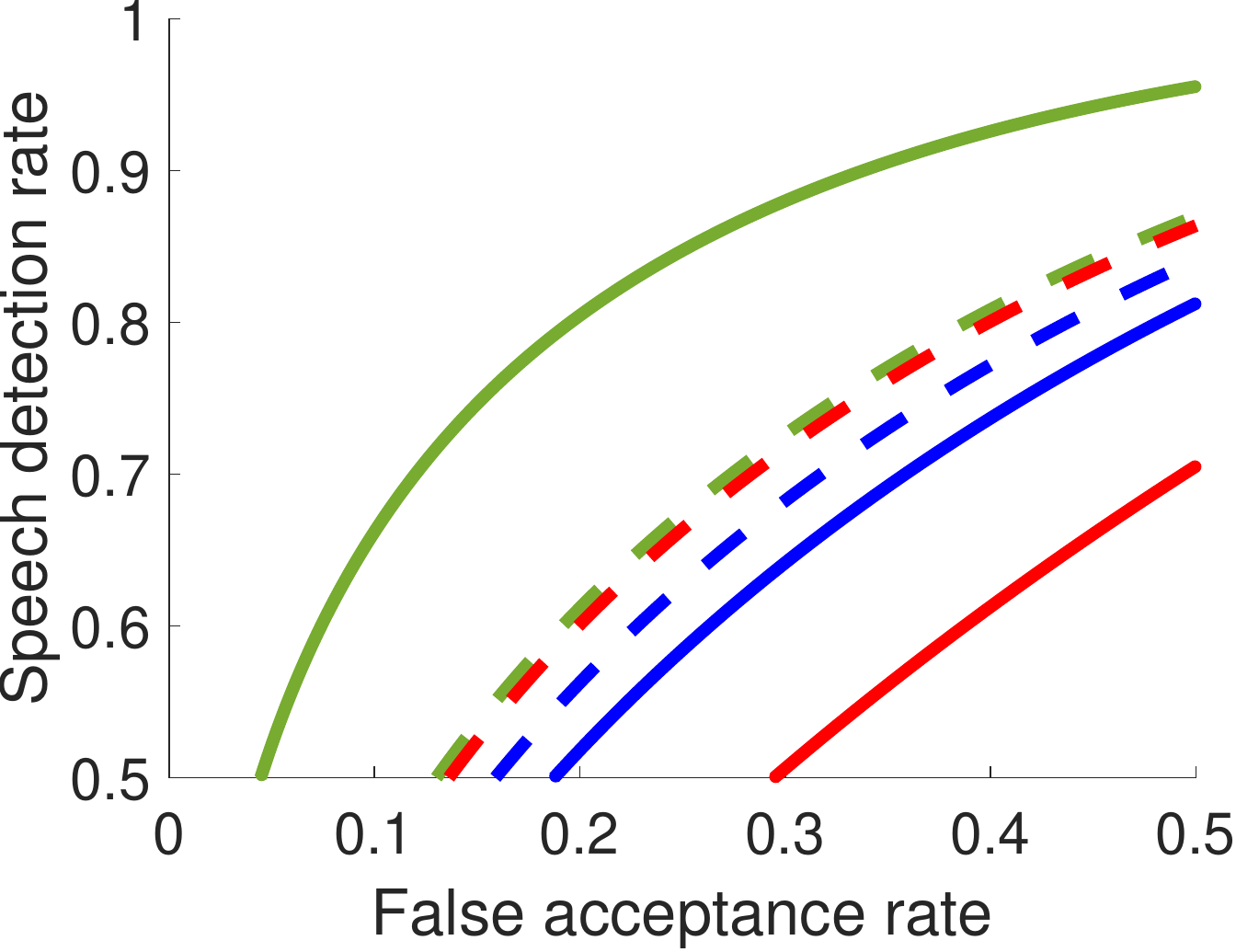}}
\captionsetup{belowskip=-10pt}
\caption{Comparison of ROC curves for the two spatial detector approaches combined with the three SVAD algorithms}
\label{fig:results_combine}
\end{figure}

The ROC curves for the first evaluation can be seen in Figure~\ref{fig:results_combine} for white and babble noise at \SIlist{0;10}{\decibel} SNR.
The AUC for both noise types and four SNRs are listed in Table~\ref{tab:table1}.
From the results it is observed that for the rVAD it is best to use the filter method, whereas for the G729B and SOHN VAD it is best to use the spatial VAD method.
\begin{table}[t!]
\renewcommand{\arraystretch}{1.1}
\caption{Comparison of AUC for the two detector approaches for the three SVAD algorithms. The best results for each SVAD are highlighted in bold}\label{tab:table1}
\resizebox{\linewidth}{!}{\begin{tabular}{@{}ccccccc@{}}\toprule
\multirow{2}{*}{\begin{tabular}{l} SNR \\ (dB) \end{tabular}} & \multicolumn{2}{c}{rVAD} & \multicolumn{2}{c}{SOHN} & \multicolumn{2}{c}{G729B}  \\ \cmidrule(lr){2-3}\cmidrule(lr){4-5} \cmidrule(lr){6-7}
 & F & A & F & A & F & A \\ \midrule
 \multicolumn{2}{l}{White noise} &&&&& \\ [0.2cm]
 -\,5 & \textbf{0.938} & 0.749 & 0.554 & \textbf{0.780} & 0.742 & \textbf{0.798}\\
 \phantom{1}0 & \textbf{0.966} & 0.827 & 0.587 & \textbf{0.873}  & 0.809 & \textbf{0.914}\\
 10 & \textbf{0.971} & 0.756 & 0.633 & \textbf{0.938} & 0.897 & \textbf{0.965}\\
 20 & \textbf{0.967} & 0.660 & 0.685 & \textbf{0.961} & 0.980 & \textbf{0.985}\\
 avg & \textbf{0.960} & 0.748 & 0.615 & \textbf{0.888} & 0.857 & \textbf{0.915}\\\Tstrut \\ [-0.4cm]
 \multicolumn{2}{l}{Babble noise} &&&&& \\ [0.2cm]
 -\,5 & \textbf{0.873} & 0.726 & 0.613 & \textbf{0.720} & 0.683 & \textbf{0.699}\\
 \phantom{1}0 & \textbf{0.885} & 0.787 & 0.649 & \textbf{0.781} & 0.735 & \textbf{0.760}\\
 10 & \textbf{0.950} & 0.901 & 0.714 & \textbf{0.858} & 0.843 & \textbf{0.849}\\
 20 & \textbf{0.959} & 0.954 & 0.780 & \textbf{0.907} & \textbf{0.950} & 0.938\\
 avg & \textbf{0.917} & 0.842 & 0.689 & \textbf{0.817} & 0.803 & \textbf{0.812} \Tstrut \\
\bottomrule
\end{tabular}}
\end{table}
\subsection{Evaluation of the SVAD algorithms with pre-processing}
\begin{figure}[H]
    \centering
    \subfloat[rVAD at \SI{10}{\decibel} \label{subfig:rvad_10db}]
    {\includegraphics[width=0.5\linewidth]{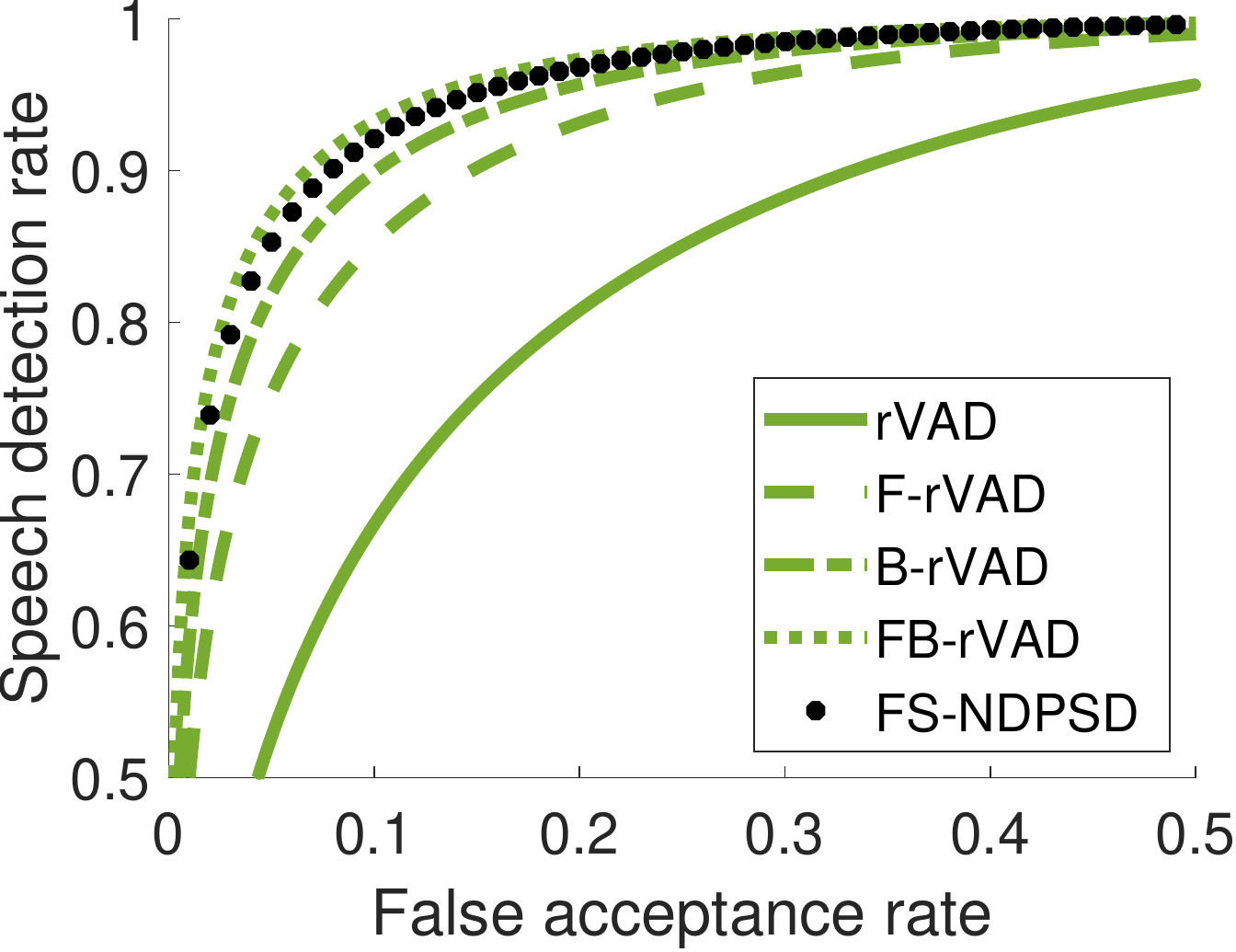}}
    \subfloat[rVAD at \SI{0}{\decibel}\label{subfig:rvad_0db}]
    {\includegraphics[width=0.5\linewidth]{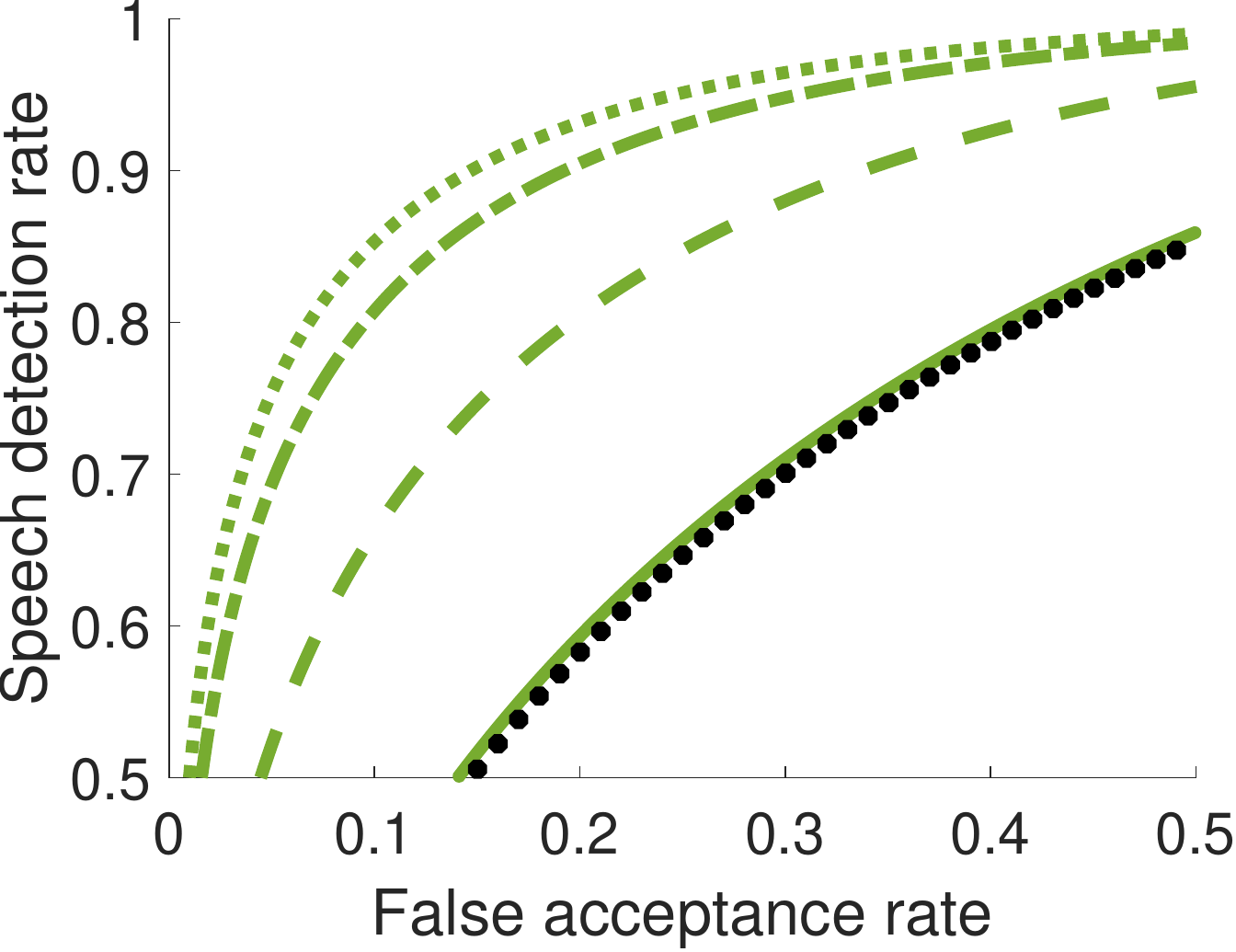}} \,
    \subfloat[G729B at \SI{10}{\decibel}\label{subfig:g729_10db}]
    {\includegraphics[width=0.5\linewidth]{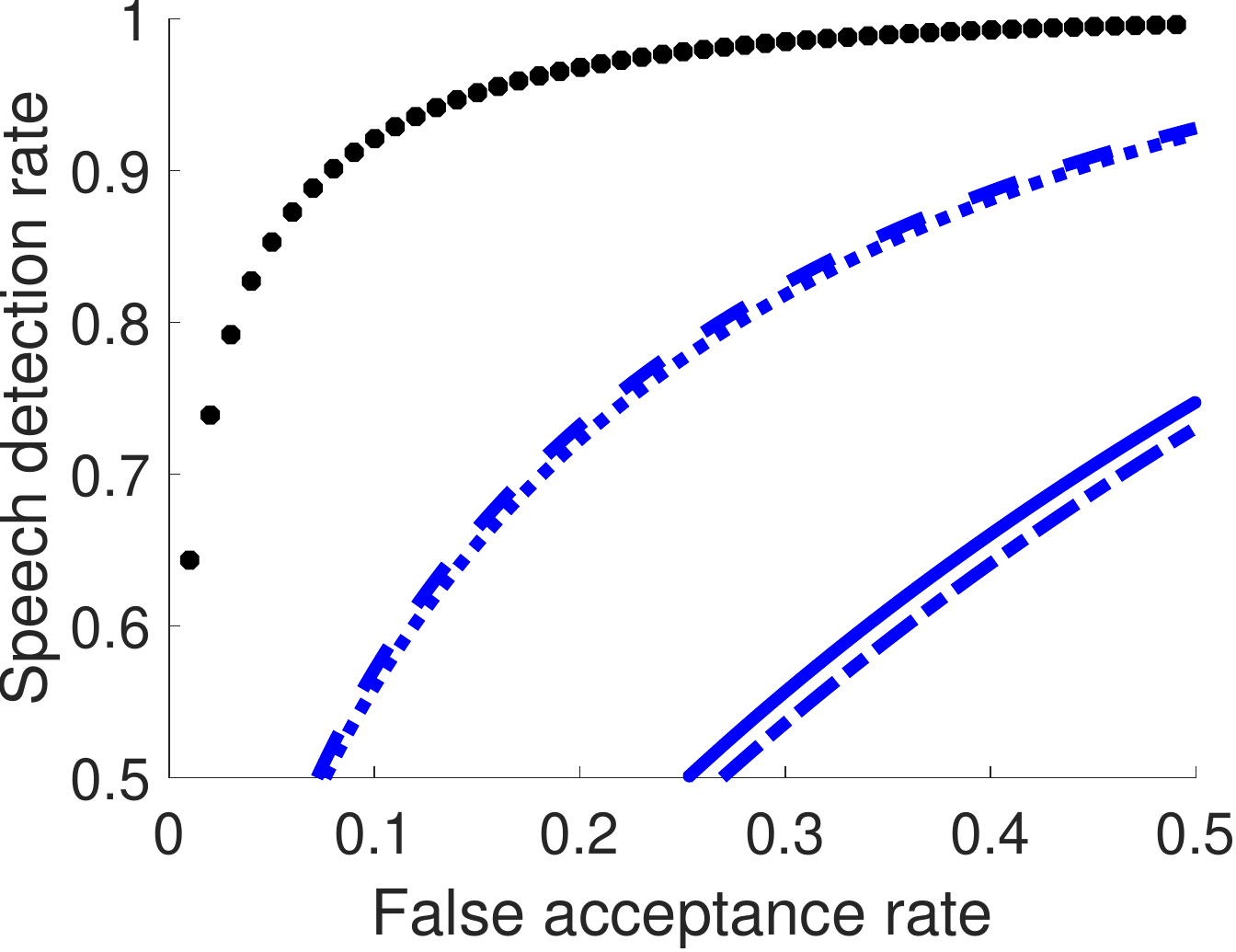}}
    \subfloat[G729B at \SI{0}{\decibel}\label{subfig:g729_0db}]
    {\includegraphics[width=0.5\linewidth]{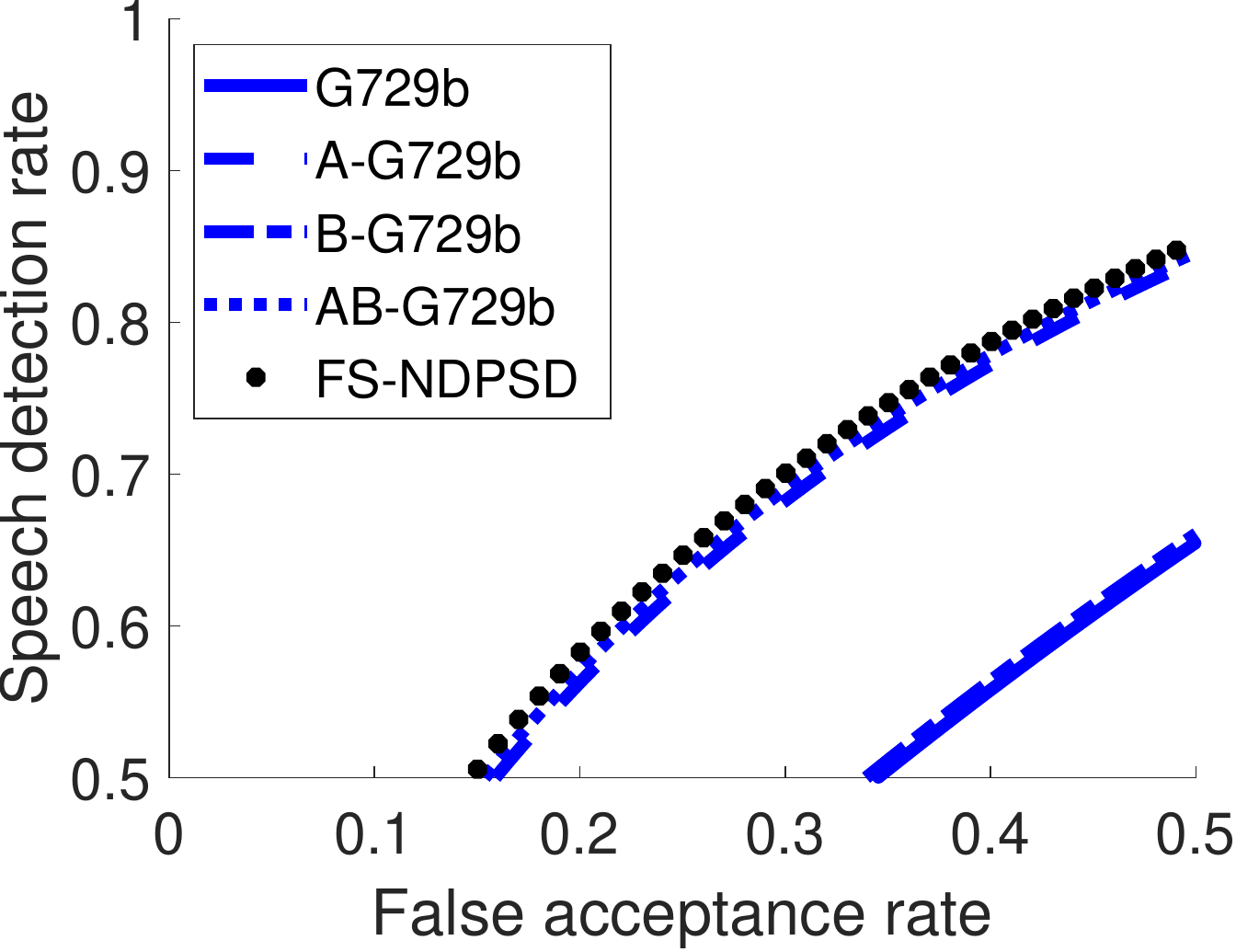}} \,
    \subfloat[SOHN at \SI{10}{\decibel}\label{subfig:sohn_10db}]
    {\includegraphics[width=0.5\linewidth]{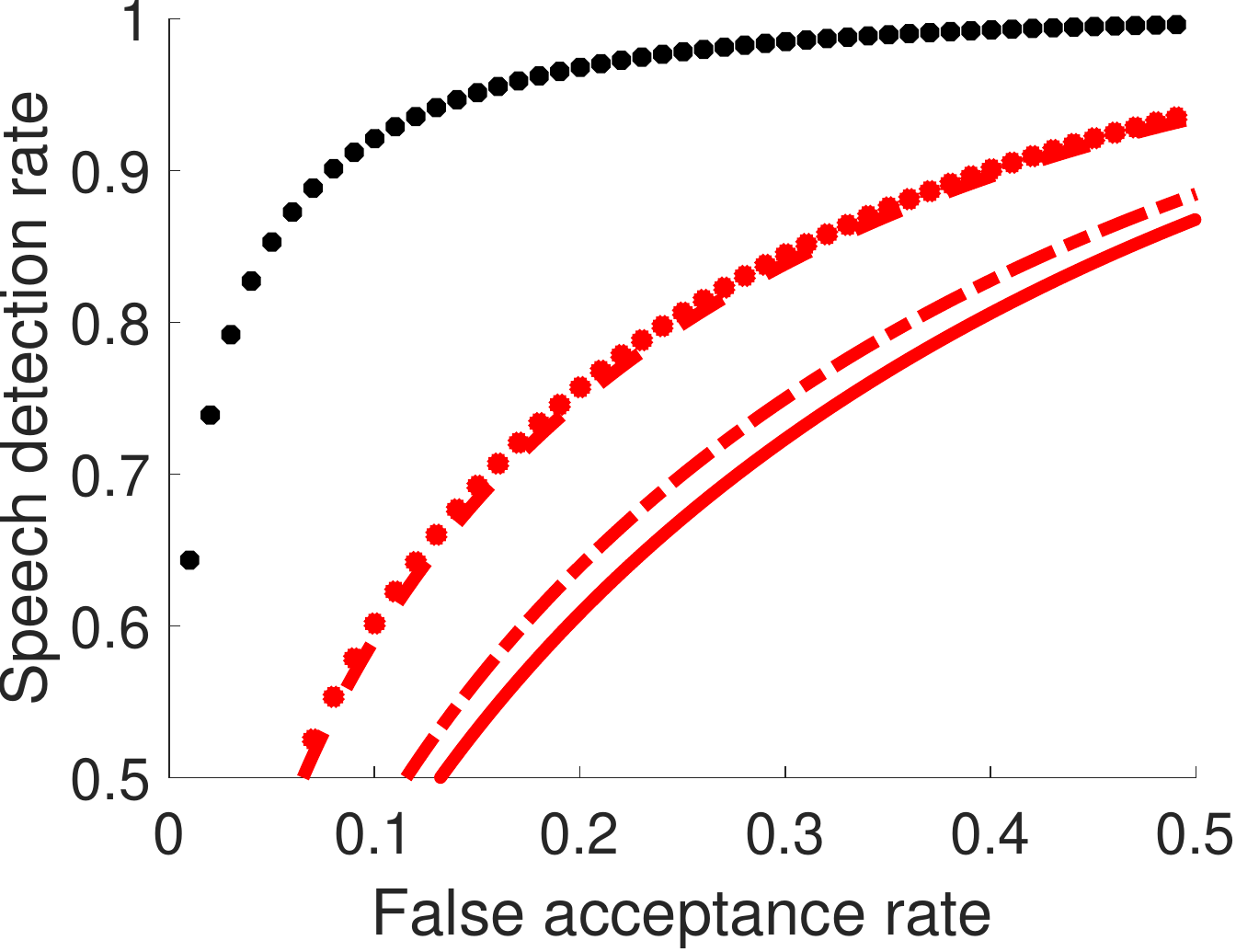}}
    \subfloat[SOHN at \SI{0}{\decibel}\label{subfig:sohn_0db}]
    {\includegraphics[width=0.5\linewidth]{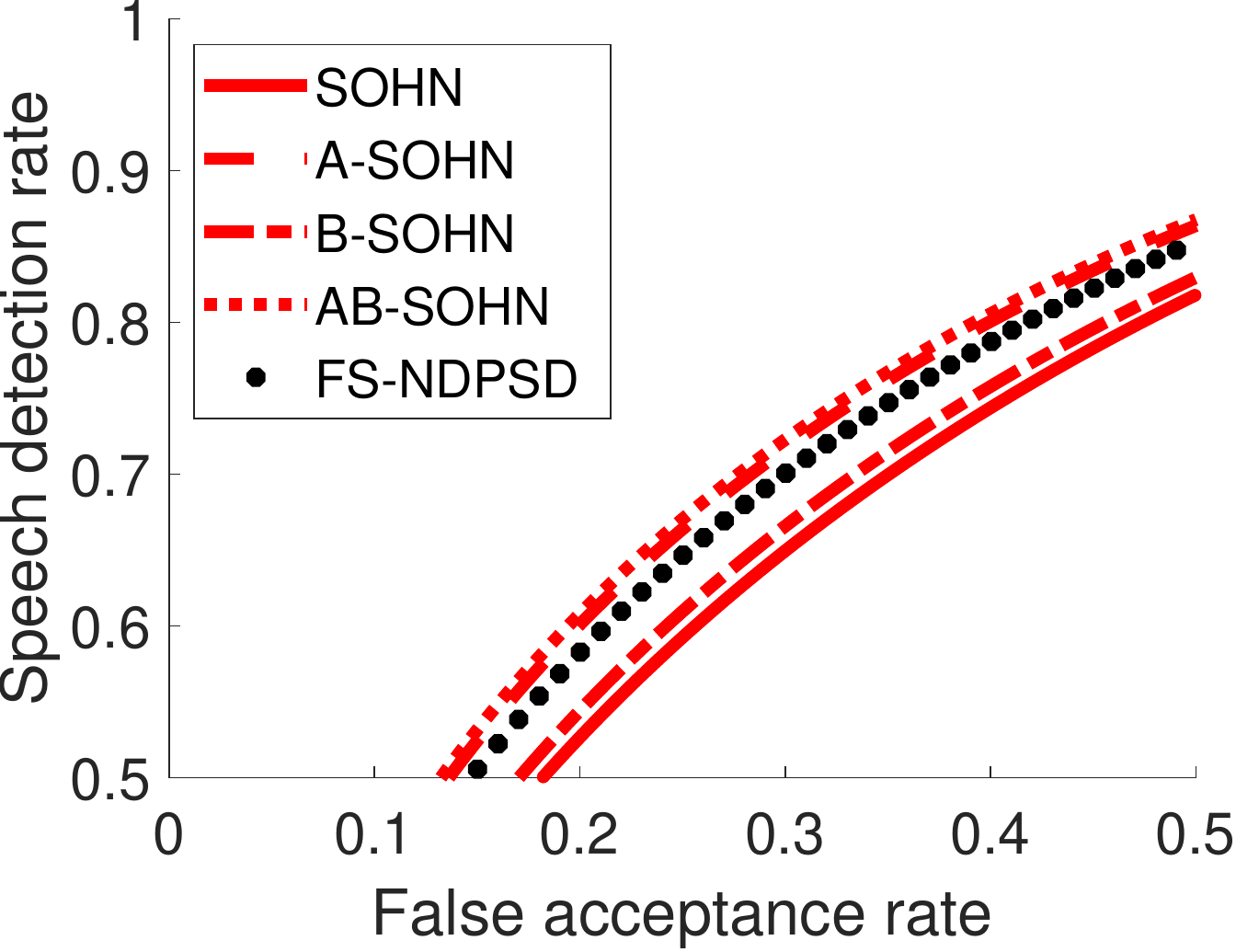}} \,
    \caption{ROC curves with comparison of SVAD, SVAD with pre-processing and FS-NDPSD MVAD, all evaluated on babble noise at the listed SNR}
    \label{fig:results}
\end{figure}

The results of the second evaluation test can be seen in Figure~\ref{fig:results} for babble noise at \SIlist{0;10}{\decibel} SNR.
The AUC for both noise types and every SNR are listed in Table~\ref{tab:table2}.
The results show that in almost every test case, the presented pre-processing methods improve the performance of the SVAD. 
In most cases the best results is the combination of the best spatial detector approach and DS with the SVAD.
However, in two instances; both noise types for SOHN and babble noise for G729B, the A-SOHN and A-G729B perform similarly to the AB-SOHN and AB-G729B respectively.
\\
When comparing the FS-NDPSD with the other evaluated VAD algorithms, the results are as follows:
For all cases where the SNR is \SIlist{-5;0}{\decibel} the FS-NDPSD performs worse than any SVAD using the best pre-processing method except for G729B in \SI{0}{\decibel} babble noise.
On the contrary, for all cases where the SNR is \SIlist{10;20}{\decibel} the FS-NDPSD is better except for FB-rVAD in \SI{10}{\decibel} babble noise.
\begin{table*}[ht]
\renewcommand{\arraystretch}{1.1}
\caption{Comparison of AUC for the four VAD algorithms including pre-processing. The best results for each SVAD algorithm are highlighted in bold. The best result for each SNR value is highlighted with underline.}
\label{tab:table2}
\resizebox{\textwidth}{!}{\begin{tabular}{@{}cccccccccccccc@{}}\toprule
\multirow{2}{*}{\begin{tabular}{c} SNR \\ (dB) \end{tabular}} & \multicolumn{4}{c}{rVAD} & \multicolumn{4}{c}{SOHN} & \multicolumn{4}{c}{G729B} & FS-NDPSD \\ \cmidrule(lr){2-5}\cmidrule(lr){6-9} \cmidrule(lr){10-13}
 & Single & B & F & FB & Single & B & A & AB & Single & B & A & AB & (MVAD) \\ \midrule
 White noise &&&&&&&&&&&&& \\ [0.2cm]
 -\,5 & 0.845 & \underline{\textbf{0.964}} & 0.938 & 0.955 & 0.614 & 0.667 & 0.780 & \textbf{0.802} & 0.638 & 0.729 & 0.798 & \textbf{0.851} & 0.599 \\
 \phantom{1}0 & 0.941 & 0.972 & 0.966 & \underline{\textbf{0.975}} & 0.697 & 0.734 & 0.873 & \textbf{0.881} & 0.775 & 0.820 & 0.914 & \textbf{0.926} & 0.786 \\
 10 & 0.962 & 0.976 & 0.971 & \textbf{0.979} & 0.784 & 0.798 & \textbf{0.938} & \textbf{0.938} & 0.865 & 0.877 & \textbf{0.965} & \textbf{0.965} & \underline{0.997} \\
 20 & 0.965 & 0.977 & 0.967 & \textbf{0.978} & 0.835 & 0.848 & \textbf{0.961} & \textbf{0.961} & 0.935 & 0.928 & \textbf{0.985} & 0.981 & \underline{0.999} \\ 
 avg & 0.928 & \underline{\textbf{0.972}} & 0.960 & \underline{\textbf{0.972}} & 0.732 & 0.762 & 0.888 & \textbf{0.896} & 0.803 & 0.838 & 0.915 & \textbf{0.931} & 0.845\Tstrut  \\ [0.2cm] %
 Babble noise &&&&&&&&&&&&& \\ [0.2cm]
 -\,5 & 0.850 & 0.919 & 0.873 & \underline{\textbf{0.922}} & 0.712 & 0.718 & 0.720 & \textbf{0.724} & 0.592 & 0.583 & 0.699 & \textbf{0.702} & 0.639   \\
  \phantom{1}0 & 0.777 & 0.936 & 0.885 & \underline{\textbf{0.950}} & 0.740 & 0.750 & 0.781 & \textbf{0.785} & 0.611 & 0.615 & 0.760 & \textbf{0.765} & 0.771 \\
  10 & 0.887 & 0.965 & 0.950 & \underline{\textbf{0.975}} & 0.785 & 0.802 & 0.858 & \textbf{0.862} & 0.681 & 0.668 & \textbf{0.849} & 0.844 & 0.971 \\
  20 & 0.935 & 0.968 & 0.959 & \textbf{0.975} & 0.823 & 0.841 & 0.907 & \textbf{0.911} & 0.855 & 0.823 & \textbf{0.938} & 0.924 & \underline{0.998} \\ 
 avg & 0.862 & 0.947 & 0.917 & \underline{\textbf{0.955}} & 0.765 & 0.778 & 0.817 & \textbf{0.820} & 0.685 & 0.672 & \textbf{0.812} & 0.809 & 0.845 \Tstrut \\
\bottomrule
\end{tabular}}
\setlength{\belowcaptionskip}{0pt}
\end{table*}

\subsection{Discussion} \label{sec:discussion}
The simulated experiment yields the best possible test scenario for the DS beamformer.
This is because the target is located directly in front of the center of the two microphones where the ITD is equal to zero, and the target location is constant.
If the target were to be moved in either direction, this beamformer would not perform as well as the proposed method.
The FS-NDPSD is originally designed for a mobile phone handset position and evaluated on a different experimental setup in \cite[p. 5-6]{dualMicReliableSpatialCues}.
Therefore, the results for the FS-NDPSD in this paper are not comparable with those obtained in the original paper.

\section{Conclusions} \label{sec:conclusion}
This paper presented two methods to improve single-channel VAD (SVAD) algorithms.
The first method is a spatial target detector, which sets signal frames to zero if the interchannel time difference of the frame is not within two predefined thresholds.
The second method is to use a beamformer technique as a pre-processor; a delay-and-sum (DS) beamformer is chosen in this study, but any other beamforming technique can be used.
The spatial target detector is used either to filter the signal sent to the SVAD, or as a spatial VAD that is combined with the SVAD decision via an AND operation.
The beamforming method can be combined with the SVAD algorithms alone or additionally combined with the two spatial detector approaches.

Based on the results, it is concluded that the performance of SVAD algorithms can be significantly improved by applying the presented pre-processing methods across all signal-to-noise ratios (SNRs).
In most cases, the spatial detector methods outperforms the DS beamformer, but the best performance is obtained from a combination of the two pre-processing methods.
At \SIlist{-5;0}{\decibel} SNR the single-channel VAD algorithms with pre-processing perform significantly better than a baseline multi-channel VAD algorithm.
\bibliographystyle{IEEEtran}
\bibliography{article}
\end{document}